\documentclass[twocolumn,A4]{aa}
\usepackage{graphicx}
\usepackage{txfonts}
\begin{document}

   \title{Mapping the cold molecular gas in a cooling flow cluster:\\
Abell 1795}

   \author{P. Salom\'e
          \inst{1}
          \and
          F. Combes\inst{1}
          }

   \offprints{P. Salom\'e}

   \institute{LERMA, Observatoire de Paris, 61, Av. de l'Observatoire
              75014 Paris\\
              \email{philippe.salome@bspm.fr}
              }
   \date{Received; accepted}

   \abstract{
Cold molecular gas is found in several clusters of galaxies (Edge,
2001, Salom\'e \& Combes, 2003): single dish telescope observations in
CO(1-0) and CO(2-1) emission lines have revealed the existence of
large amounts of cold gas (up to $\sim$10$^{11}$M$_\odot$) in the
central region of cooling flow clusters. We present here
interferometric observations performed with the IRAM Plateau de Bure
interferometer in Abell 1795. Comparison with IRAM 30m data shows the
cold gas detected is extended suggesting a cooling flow origin.  
The CO features identified are very similar to the
structures observed in H$_\alpha$ and with the star forming regions observed
through UV continuum excess. A large fraction of the cold gas is not
centered on the central cD, but located near 
brightest X-ray emitting regions along the North-West orientated radio
lobe. The cold gas kinematics is consistent with the optical
nebulosity behaviour in the very central
region. It is not in rotation
around the central cD : a velocity gradient shows the cold gas 
might be cooled gas from the intra-cluster medium being accreted by the central galaxy. 
The optical filaments, aligned with the cD orbit, are intimately related to the
radio jets and lobes. The material fueling the star formation
certainly comes from the deposited gas, cooling more
efficiently along the edge of the radio lobes. Even 
if some heating mechanisms are present, these millimetric observations show
that an effective cooling to very low temperatures indeed occurs and is probably
accelerated by the presence of the radio source. 


}
   \maketitle

\section{Introduction}

\subsection{Abell 1795}

 The rich cluster of galaxies Abell 1795 is thought to be the site of
 a cooling flow, since the temperature measured from X-ray emission is
 dropping by a factor 3 towards the core, at about 140 kpc from the
 centre (e.g. Tamura et al 2001). Fabian et al (2001) discovered with
 the Chandra satellite a $40''$ long ($\sim$60 kpc, with an adopted
 Hubble constant of H$_0$ = 75 km/s/Mpc) X-ray filament in its core,
 coinciding with an H$_\alpha$+NII filament previously found by Cowie et
 al. (1983). This filament is also conspicuous in the U-band (McNamara
 et al 1996a), and the site of star formation: the blue continuum
 along the edges of the radio lobes are resolved into bright knots
 with the Hubble Space Telescope (HST). The absence of polarized light
 rules out that the U-band continuum is only due to scattered light
 from an active nucleus (McNamara et al. 1996b). The cD galaxy, a
 little South of the Northern peak of the filament, has a positive
 peculiar velocity with respect to the rest of the cluster galaxies
 (Oegerle \& Hill 1994), and is probably oscillating around the cluster
 core. The gas filament has a velocity centred on the cluster mean
 velocity, and may also be sloshing in the cluster potential
 (Markevitch et al. 2001). The radiative cooling time of the X-ray
 emitting gas in the filament is about $3\times 10^8$ yr, quite similar to the
 dynamical age of the filament (ratio of length to velocity). This
 supports the cooling flow model, in which the gas is presently
 cooling from the hot cluster gas. The optical filaments are
 intimately related to the radio jets and
lobes from the radiosource 4C+26.42 (vanBreugel et al. 1984, Ge \& Owen
1993, Mc Namara et al 1996). The blue emission is aligned along the
edges of the radio lobes, while dust lanes extend along the radio jets
and lobes as H$_\alpha$ and UV (Pinkney et al 1996), which might be
interpreted as star formation induced by the radio source or the
deflection of the radio jets off pre-existing dust and gas. If a burst
of star formation were triggered by the expanding radio lobes, the age
of the burst population should be $\sim$$10^7$ yr. Then, the star formation
rate in both lobes, assuming the local IMF, would be $\sim$20 M$_\odot$/yr, and
the stellar mass of the lobes would be $\sim$$10^8$ M$_\odot$ (McNamara et
al. 1996a). The material fuelling the star formation and the radio
source may have two origins: either the cooling flow or gas tidally
 stripped by galatic interactions (dense molecular clouds are not sensitive to ram
 pressure). The morphology and kinematics of CO emission
could distinguish between the two possibilities. 

\subsection{Cold molecular gas}

Previous spatially resolved CO emission in a cooling flow cluster has
already been reported in the Perseus galaxy NGC 1275, with a
morphology related to H$_\alpha$ and X-ray (e.g. Inoue et
al. (1996), Bridges \& Irwin (1998), using the Nobeyama Millimeter Array
and the IRAM 30m single dish telescope respectively). However,
this galaxy is also the result of a merger, the CO is observed in the
center in rotation around the AGN, and the origin of the CO gas is
multiple, including the cooling flow. The picture appears clearer in
Abell 1795. Recent OVRO observations, (Edge \& Frayer, 2003) observed 
CO emission in a compact region centered on the central galaxy in
five cooling clusters (Abell 1068, RXJ0821+07, Zw3146, Abell 1835 and
RXJ0338+09). 

The core of the Abell 1795 has also been observed in the CO rotational line
emission, tracing the cold molecular hydrogen. The total emission was
first detected with the IRAM 30m telescope (Salom\'e \& Combes
2003) and the mass of the cold gas in a region of 23$"$ was estimated to
be $4.8\times 10^9$ M$_\odot$. It is likely to be a lower 
limit, since the measured metallicity
of the Intra-Cluster Medium (ICM) is 0.2-0.3 solar (Tamura et
al. 2001). And much more gas might be present into the cooling
radius. The molecular mass would then be much larger, by an order of
magnitude, than in a typical cD galaxy. The mass deposition rates
deduced from recent Chandra X-ray observation of Abell 1795 are
$\sim$7.9 M$_\odot$/yr in the central region covered by the 30m telescope beam of 22$"$
(while it is $\sim$100 M$_\odot$/yr within 200 kpc, Ettori et al 2002), therefore
compatible with our detection of cold molecular gas. The cold gas
detected might have been deposited out of the flow in $\sim$600 Myr. The
cooling time of the hot X-ray gas is around $\sim$300 Myr in the same
region. In a steady state cooling flow scenario, a steady reservoir of
cold gas with a mass close to the mass deduced from CO observations is
possible since the star formation rate (Smith et al. 1997, McNamara et
al. 1996b) is close to the mass deposition rate. 
However, the lack of spatial resolution prevented to
conclude that the cold gas was associated to the cooling flow, since
it could have been interstellar gas rotating in the central cD
galaxy. The gas detected here through CO rotational lines is cold (about
20K). The intensity ratio between the CO(1-0) and CO(2-1) (obtained
with the IRAM-30m, Salom\'e© \& Combes 2003) is consistent with an
optically thick gas.
Warm H$_2$, vibrationnally excited, is particularly
abundant in cooling flow galaxies and in Abell 1795 (Donahue et al
2000, Wilman et al. 2002). The warm H$_2$ emission could be related to
the interaction between the jets and the cold molecular gas.


\section{Interferometric observations}

The observations we present here, have been done with the IRAM
interferometer in winter 2003, with 3.2$"$ and 1.8$"$ spatial resolution at the CO(1-0)
and CO(2-1) lines respectively. The frequencies were centred at
108.413 and 216.822 Ghz, corresponding to the redshift $z=0.06326$ of
the cluster cD galaxy hosting the 4C+26.42 radiosource. The velocity
of the cD is redshifted by 374 km/s with respect to the mean velocity
of the galaxies inside 200 kpc (Hill et al. 1988, Oegerle \& Hill 1994).
The total integration time is 43 hours, in C and D configuration, with
5 or 6 antennas. To improve the signal-to-noise ratio, we smoothed the
spectral resolution to channels of 88 km/s at CO(1-0) and 44 km/s at
CO(2-1), since the width of the line is expected to be large ($\sim$500
km/s) from the 30m observations. The signal-to-noise ratio is slightly
better in the CO(2-1) map, and the higher spatial resolution allows to
better identify the H$_\alpha$/CO correspondance; however, the primary
beam (FWHP=22$"$) is twice smaller than in CO(1-0) (FWHP=45$"$),
preventing to observe the spatial extension of the filaments .


\section{Results}


       \subsection{The molecular gas morphology}

\begin{figure*}[htbp]
\begin{tabular}{c}
\includegraphics[width=15cm]{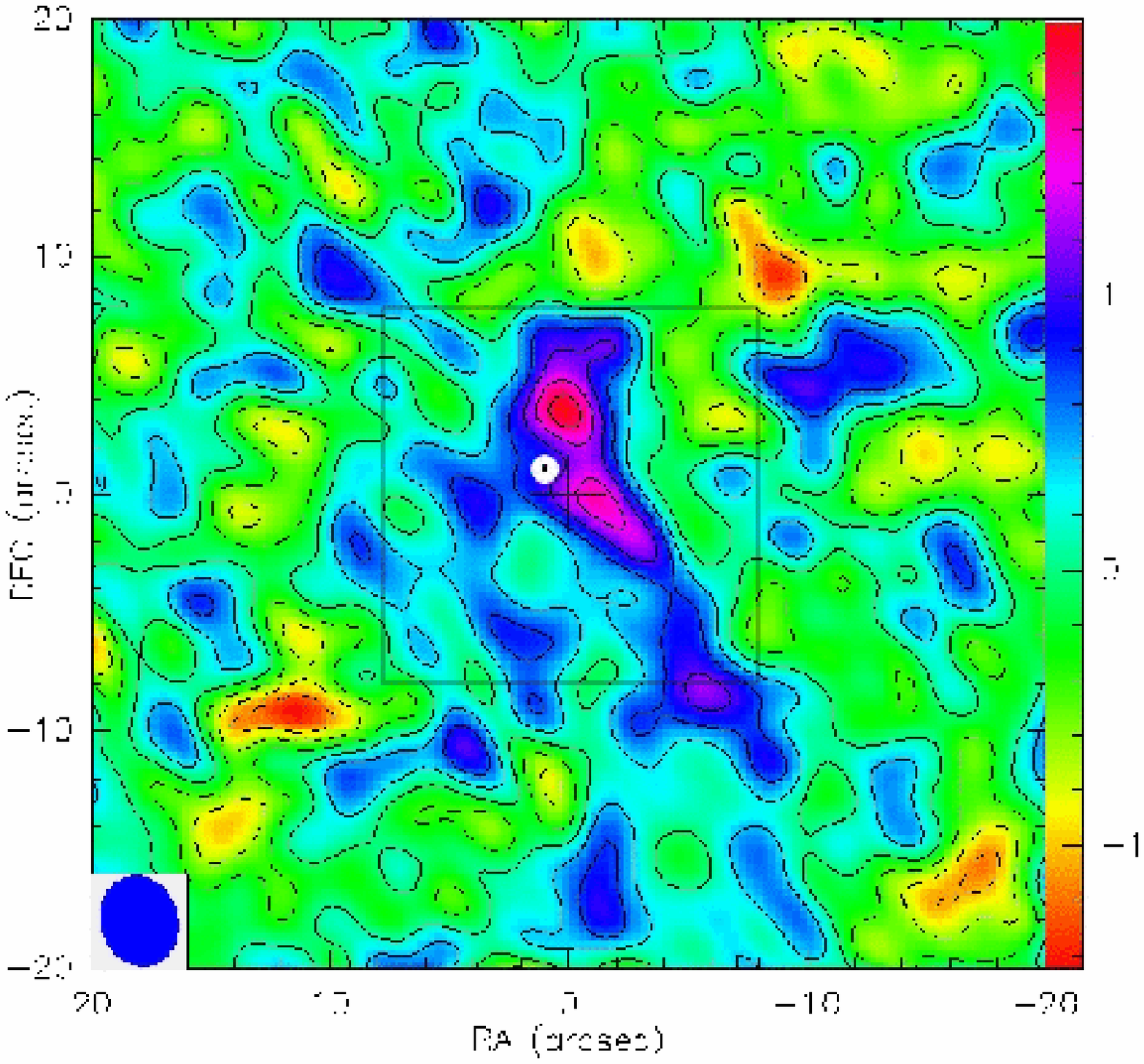}\\
\includegraphics[width=19cm]{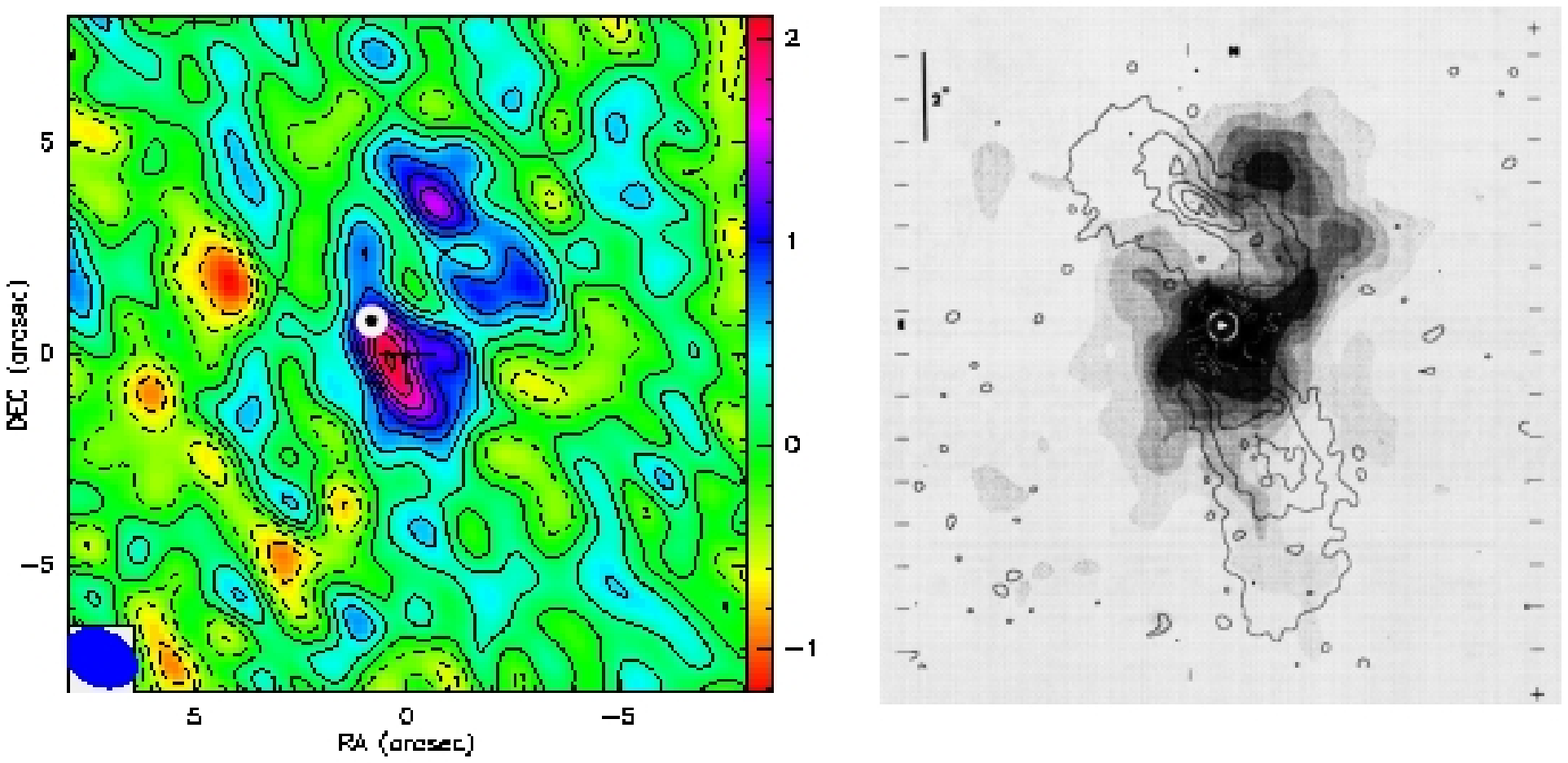}\\
\end{tabular}
\caption{
Top : C0(1-0) integrated emission. Linear contours are drawn from
$-$3$\sigma$ to 6$\sigma$ spaced by 1$\sigma$ = 0.36 Jy/beam.km/s. Dashed
are negative contours. The beam is plotted in the bottom left
corner. The black box indicate the size of the CO(2-1) integrated map
presented in the bottom left. Contours are from
$-$3$\sigma$ to 8$\sigma$ spaced by 1$\sigma$ = 0.26 Jy/beam.km/s. The
white disk indicate the radio source position. Bottom right :
H$_\alpha$ + [NII] line emission in grey scale, overlaid the 6cm
continuum emission from 4C+26.42 radio lobes (vanBreugel et al. 1984).  }
\label{maps}
\end{figure*}

The resulting integrated maps in the two CO lines are shown in Figure
\ref{maps}. At 3mm, we detect a continuum source, at the position of
4C+26.42, of 7 mJy. This source is the expected continuation of the
synchrotron emission detected at lower frequency, with a flux
decreasing slope of $\alpha$ = $-$0.98. The CO(1-0) map shows the emission found,
once the continuum has been subtracted. We clearly detect CO emission
associated with the cooling region already detected in X-rays, U-band
excess and H$_\alpha$. The maximum of the emission is located in two main
regions : one coincident with the maximum of X-ray emission and occurs
at the North-West of the cD, the other is at the galaxy position and
extending to the South. The CO emission is too faint to be compared to
the large North-South orientated filamentary structure seen in X-ray
and H$_\alpha$. Nevertheless, in the very central part, the cold gas
morphology is very similar to the H$_\alpha$ and Blue continuum structures
identified by Van Breugel et al. (1984), McNamara \& O'Connell (1993),
Smith et al (1997) as shown on Fig. \ref{maps}. The H$_\alpha$-CO correlation has
been found for global emission statistically over the CO-detected
cooling flows (Edge 2001, Salom\'e© \& Combes 2003), and it is now
confirmed by their coinciding morphologies in A1795.

     \subsection{The molecular gas kinematics}

\begin{figure}[htbp]
\vspace{-2.8cm}\includegraphics[width=8.5cm,angle=-90]{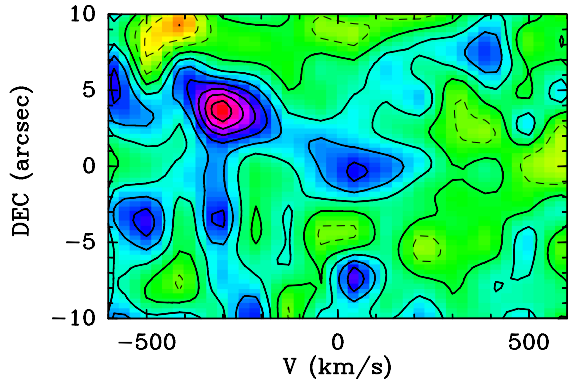}
\vspace{-1.cm}\caption{Position-Velocity diagram in CO(1-0) emission line, the 
positions are along a slit of 5$\arcsec$ width (integrated),
centered on the galaxy position and aligned with the maximum of
emission (PA=27°).
\label{pv10}}
\end{figure}

\begin{figure}[htbp]
\vspace{-3.4cm}\includegraphics[width=8.5cm,angle=-90]{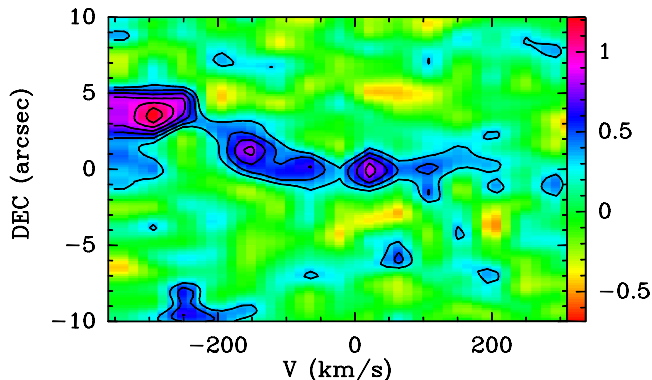}
\vspace{-1.1cm}\caption{Position-velocity map in CO(2-1) emission line in the same 
region as in Fig. \ref{pv10}.
\label{pv21}}
\end{figure}

The kinematics of the molecular gas in A1795 is shown on Figures
\ref{pv10} and \ref{pv21}. We present position-velocity diagrams for
the two CO lines along the axis of maximum emission (PA=27°) through
the centre of the cD galaxy. There are two separated trends in
velocity in both the CO(1-0) and CO(2-1) lines. 
One is centered on the galaxy position at zero velocity. The other  
is 5'' North-West from the radio source position and is at $-$300/350 km/s 
relative to the galaxy velocity. There is also a regular
gradient from the galaxy centre towards the North-West (5''). 
The velocities
measured in the CO brightest regions are coinciding with those in
H$_\alpha$, by Van Breugel et al (1984) and Anton (1993), supporting the association
between the cold gas and the hot gas/optical structures.  The peak of
H$_\alpha$ emission does not follow the peculiar velocity of the cD galaxy
in the cluster, but is centred on the cluster mean velocity
($-$350km/s), i.e. the mean velocity of the galaxies inside 200 kpc,
while it reaches the cD velocity at the galaxy position. The
kinematics of the cold gas and the H$_\alpha$ line emitting gas is
compatible with the cooling flow scenario. At the North-West position,
coincident with the brigthest X-ray region, the gas is cooling within
the cluster potential. Near the cD, the flow is captured by the cD
potential (and the inner gas have a velocity close to that of the
galaxy, with a wider range of velocities).

\vspace{-0.7cm}

\section{Discussion}

The peculiar morphology of the cooling gas (X-ray, H$_\alpha$, blue
continuum, molecular gas) appears to avoid the radio lobes,
particularly visible in the north (Fig. \ref{maps}). This can be interpreted as
the radio jet heating and displacing the hot X-ray gas as it cools on
the cD. This scenario is supported by the detection of X-ray cavities
coincident with radio lobes in an increasing number of cooling flow
clusters imaged by the recent high spatial resolution X-ray satellite
Chandra (B\"ohringer et al., 1993; Fabian et al., 2000). The cold
molecular gas will condense more efficiently in the resulting
cold front surrounding the cavities. This is also in these fronts
that stars are formed, explaining the optical blue continuum. The CO
emission is not associated to the nearby galaxies of the cluster. It
cannot be all tidally stripped gas, since most of 
the molecular gas detected is not centered on the galaxy, see Fig. 2
and 3. The association of the CO
emission with the cavity border supports its origin in an intermittent cooling
flow scenario, where the gas cooling more rapidly along the radio
lobes is then accreted by the central galaxy and can fuel the central AGN
activity, which may regulate the cooling. Finally, an important result
of these interferometric observations is that the
total flux retrieved is only 25\% of the single dish flux obtained
previously with the IRAM 30m telescope: this means that most of the
emission is extended with respect to the 45$"$ beam (corresponding to
65 kpc), which argues for a cooling flow origin of the cold gas at
larger radii.

\vspace{-0.3cm}
\section{Conclusions}

Through interferometric mapping of the CO emission, we have shown that
the cold molecular gas is associated to the cooling flow in Abell
1795. The CO emission is closely associated to H$\alpha$ and to X-ray
emissions, and is concentrated at the boundaries of the bubbles or
cavities created by the central AGN. Cooling occurs preferentially at
the edge of these cavities, were the hot gas is denser. The peculiar
long filament morphology of the cooling gas in A1795, and its
kinematics, are best interpreted as a cooling wake (Fabian et
al. 2001): the cD galaxy oscillates in a few $10^8$ yr period, and
during a cooling time the large-scale hot gas sees the minimum of the
potential roughly as a straight line along the cD orbit. The AGN in
the cD core and its plasma jets certainly provide a feedback
heating, seen as cavities in the X-ray maps, coincident with the
radio-lobes, but is also probably increasing the cooling to very low
temperature along the edges of these cavities, where the cold gas
condenses and forms stars. A velocity gradient of the cold gas is
revealed, that shows it is falling on the central galaxy and
may provide the AGN fueling material which is consistent with an AGN
regulated cooling flow scenario.
\vspace{-0.1cm}
\begin{acknowledgements}
We thank the IRAM Plateau de Bure staff for their
support. IRAM is funded by the INSU/CNRS (France), the MPG (Germany) and the
IGN (Spain). We also thank the referee for his useful comments. 
\end{acknowledgements}
\vspace{-0.5cm}


\end{document}